\documentclass[
oneclomn,
showpacs,
amsmath,amssymb]{revtex4}
\usepackage{bm}
\usepackage[dvips]{graphicx}

\begin{document}

\title{Bursting transition in a linear self-exciting point process}

\author{Tomokatsu Onaga}
\email{onaga@scphys.kyoto-u.ac.jp}
\affiliation{Department of Physics, Kyoto University, Kyoto 606-8502, Japan}
\author{Shigeru Shinomoto}
\email{shinomoto@scphys.kyoto-u.ac.jp}
\affiliation{Department of Physics, Kyoto University, Kyoto 606-8502, Japan}

\date{\today}

\begin{abstract}
Self-exciting point processes describe the manner in which every event facilitates the occurrence of succeeding events. By increasing excitability, the event occurrences start to exhibit bursts even in the absence of external stimuli. We revealed that the transition is uniquely determined by the average number of events added by a single event, $1-1/\sqrt{2} \approx 0.2929$, independently of the temporal excitation profile. We further extended the theory to multi-dimensional processes, to be able to incite or inhibit bursting in networks of agents.
\end{abstract}
\pacs{89.75.Hc, 05.40.-a}

\maketitle


Irregular occurrences of events are modeled by the Poisson process such that point events are independently drawn in time at a given rate. Event occurrences that are not mutually independent may be modeled by adding a supplementary probability for event occurrence after every event~\cite{Hawkes71a}. This simple model called the Hawkes process has been widely applied to the analysis of earthquakes~\cite{Ogata88}, genome sequences~\cite{Reynaud10}, urban crime~\cite{Mohler11}, human activity~\cite{Malmgren08,Mitchell10,Masuda13}, and neuronal activity~\cite{Krumin10,Rotter13}.

\begin{figure}[ht]
\begin{center}
\includegraphics[clip,width=8.6cm]{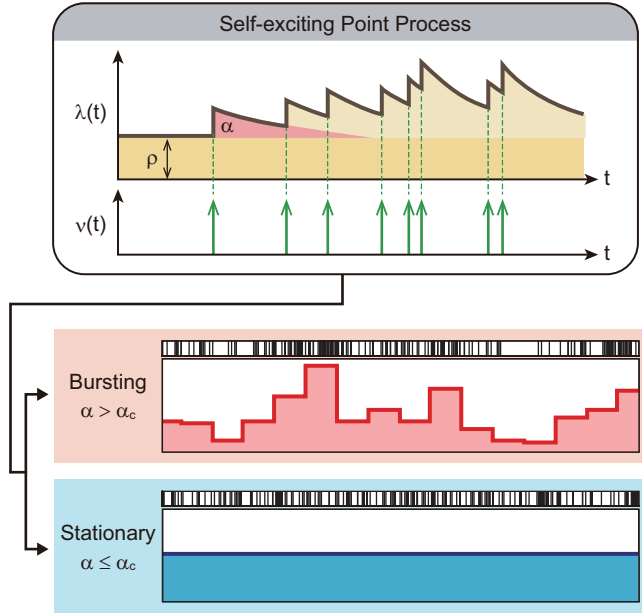}
\caption{{\bf Self-exciting processes.} (top) The rate of event occurrence $\lambda(t)$ is modulated according to generated events $\nu(t)=\sum_{k} \delta(t-t_k)$. (bottom) Event sequences depicted in rasters may exhibit non-stationary bursts or remain stationary, depending on whether the excitability $\alpha$ is larger or smaller than a critical value $\alpha_c$, respectively. The bin size of the histogram shown under each raster was selected using the method of minimizing the mean squared error between the histogram and the underlying rate.}
\label{figschematic}
\end{center}\end{figure}

The process is called self-exciting, if each event is associated with a positive supplementary probability for succeeding events, which we call ``excitability.'' Given large excitability, the system may exhibit unstable bursts of events leading to non-stationary occurrence rate, even in the absence of external stimuli (Fig.~\ref{figschematic}). Contrariwise, under small excitability, the fluctuation in the occurrence rate may become undetectable from a single sparse series of irregular events. 

A transition from undetectable to detectable fluctuation in the occurrence rate may be verified by principled rate estimators such as an optimal time histogram or the empirical Bayes rate estimator; if the rate estimators indicate a constant rate, we interpret that the fluctuation in the underlying rate is unknowable~\cite{Koyama04,Koyama07}. Herein we obtain the critical condition for the excitability at which the estimated rate changes between constant and fluctuating. Based on the second order transition, the criticality condition is obtained in a universal formula applicable to a wide range of self-exciting processes associated with various temporal excitation profiles.

We then extend our analysis to the multi-dimensional Hawkes process, in which multiple agents mutually influence each other. This process is exemplified by a social system in which people influence activity through events such as emails and web lookups~\cite{Argollo04,Rybski12}. It is known that people may exhibit autonomous bursts of activity without exogenous stimuli~\cite{Barabasi05,Sornette04,Crane08}. Knowing the conditions under which bursts occur, we can control the occurrence of burst activity by reconnecting people or agents.


In the Hawkes process, the rate of event occurrence $\lambda(t)$ is modulated by past events as
\begin{eqnarray}
\lambda(t)=\rho+\alpha\sum_{k}f(t-t_k),
\label{hawkes}
\end{eqnarray}
where $\rho$ is the base rate, $t_k$ is the occurrence time of the $k$th event. The kernel function $f(t)$, representing the time course of the supplementary probability, satisfies two conditions: the causality, $f(t)=0$ for $t<0$, and the normalization, $\int_{0}^{\infty}f(t)dt=1$. Accordingly, the coefficient $\alpha$ represents the excitability or the supplementary probability added after each event. 

Firstly, we estimate the correlation function of the event occurrence rate according to Hawkes~\cite{Hawkes71b}, and extend the originally proposed range of validity. By representing a series of event occurrences as a sum of Dirac delta functions, $\nu(t)=\sum_{k} \delta(t-t_k)$, the Hawkes process (\ref{hawkes}) may be represented as
\begin{eqnarray}
\lambda(t)=\rho+\alpha \int_{-\infty}^{\infty} f(t-u) \nu(u) du.
\label{hawkesint}
\end{eqnarray}
Because the ensemble average of the event occurrence $\langle \nu \rangle$ equals that of the rate $\langle \lambda \rangle$, the average rate is obtained as $\langle \lambda \rangle = \rho/(1-\alpha)$. The excitability $\alpha$ should be less than unity to avoid pandemic explosion, in which $\lambda(t)$ diverges.

The complete covariance density $\phi^{(c)}(s) \equiv \langle \nu(t+s) \nu(t) \rangle -\langle \lambda \rangle^2$ has a singularity at $s=0$ with $\langle \lambda \rangle \delta(s)$. Because $\langle \nu(t+s) \nu(t) \rangle = \langle \lambda(t+s) \nu(t) \rangle$ for $s > 0$, the covariance density satisfies the relation,
\begin{eqnarray}
\phi^{(c)}(s) = \alpha \int_{-\infty}^{\infty}f(s-u)\phi^{(c)}(u)du,
\label{phi_d}
\end{eqnarray}
for $s>0$. The correlation function of the rate fluctuation $\delta \lambda(t) \equiv \lambda(t) - \langle \lambda \rangle$ is given by removing the singularity from the complete covariance density, or $\phi(s) \equiv \langle \delta \lambda(t+s) \delta \lambda(t) \rangle = \phi^{(c)}(s) - \langle \lambda \rangle \delta(s)$. Inserting this relation into the integral equation (\ref{phi_d}), we obtain an integral equation for the correlation function,
\begin{equation}
\phi(s)=\alpha \langle \lambda \rangle f(s)+\alpha \int_{-\infty}^{\infty}f(s-u)\phi(u)du,
\label{integrationeqn}
\end{equation}
which holds for $s>0$. Define a function, 
\begin{equation}
g(s) \equiv \alpha \langle \lambda \rangle f(s)+\alpha \int_{-\infty}^{\infty}f(s-u)\phi(u)du-\phi(s),
\label{fung}
\end{equation}
which satisfies $g(s)=0$ for $s>0$. The Fourier transformation of this equation is
\begin{equation}
\tilde g_{\omega}=\alpha \langle \lambda \rangle\tilde f_{\omega}+\alpha \tilde f_{\omega}\tilde \phi_{\omega}-\tilde \phi_{\omega}. 
\label{gphi}
\end{equation}
Considering the time reversal symmetry of the correlation function, we obtain
\begin{equation}
\alpha \langle \lambda \rangle\tilde f_{\omega}+\left(1-\alpha \tilde f_{\omega}\right)\tilde g_{-\omega}=\alpha \langle \lambda \rangle\tilde f_{-\omega}+\left(1-\alpha \tilde f_{-\omega}\right)\tilde g_{\omega}.
\label{timereversal}
\end{equation}
Because $f(t)=0$ for $t<0$ and $g(t)=0$ for $t>0$, their Fourier images $\tilde f_{\omega}$ and $\tilde g_{\omega}$ converge $0$ in the limit of $|\omega|\to \infty$ in half planes of ${\rm Im}(\omega)<0$ and ${\rm Im}(\omega)>0$, respectively. Because the LHS and RHS of Eq.(\ref{timereversal}) are regular in the lower and upper half imaginary planes, they vanish in respective half planes. While Hawkes derived the relation assuming an exponentially decaying kernel function, we may permit long-tailed kernels, such as power law functions $f(t) \propto (1+t)^{-b}\;\left(b>1\right)$, by trimming the range of the functional regularity to the adjoining half planes of ${\rm Im}(\omega)\ge0$ and ${\rm Im}(\omega)\le0$. 

Inserting the identity relation into Eq.(\ref{gphi}), we obtain the relation
\begin{equation}
\tilde\phi_{\omega}= \frac{2\alpha \tilde f_{\omega}-\alpha ^2\tilde f_{\omega}^2}{(1-\alpha \tilde f_{\omega})^2} \langle \lambda \rangle ,
\label{fouriercorrelation}
\end{equation}
by which the correlation function $\phi(t)$ is obtained for a given excitation kernel $f(t)$.


Secondly, we derive the condition for detecting fluctuation in the rate for a single series of event times. Though the self-exciting process is a stationary process whose statistical properties are invariant with time as an ensemble, individual processes may significantly fluctuate in time, causing bursts of events. We decide the non-stationarity of a single series of events based on whether principled rate estimators indicate fluctuating rate or constant rate. Herein we construct an optimal histogram in which the bin size is selected to minimize the mean integrated squared error (MISE) between the histogram and the underlying rate, and derive the condition under which the optimal bin size diverges, or equivalently an optimal histogram indicates constant rate~\cite{Koyama04}.

The bin size $\Delta$ is selected to minimize MISE between the underlying rate $\lambda(t)$ and the histogram $\hat \lambda_{\Delta}(t)$. The MISE is a function of the bin size
\begin{equation}
S(\Delta)=\lim_{T \to \infty}\frac{1}{T}\int_0^T \left\langle  \left(\lambda(t)-\hat \lambda_{\Delta}(t)\right)^2 \right\rangle dt,
\end{equation}
where $T$ is the entire observation interval. Replacing the long time average with the average over the bin size, $\hat \lambda_{\Delta}(t)$ can be treated as a single rectangle whose height is the number of events $K$ divided by the bin size $\Delta$. Thus the MISE is given as
\begin{equation}
S(\Delta)=\left\langle \frac{1}{\Delta} \int_0^{\Delta} \left( \lambda^2(t) -\frac{2 K}{\Delta} \lambda(t) \right) dt + \frac{K^2}{\Delta^2} \right\rangle.
\label{mise1}
\end{equation}
The expected number of events in each interval is given by integrating the underlying rate: $E[K]=\int_0^{\Delta} \lambda(t) dt$. Because events are independently drawn, the Poisson relation holds: $E[K^2]=E[K]^2+E[K]$. Inserting these relations into Eq.(\ref{mise1}), we have
\begin{eqnarray}
S(\Delta)=\phi(0)+\frac{\left\langle \lambda \right\rangle}{\Delta}-\frac{1}{\Delta^2}\int_0^{\Delta}dt\int_{-t}^t \phi(s) ds,
\end{eqnarray}
where $\phi(s) = \left\langle \lambda(t+s)\lambda(t) \right\rangle-\left\langle \lambda \right\rangle^2= \left\langle \delta \lambda(t+s) \delta \lambda(t) \right\rangle$.

If a series of events is derived from a constant rate process, the MISE is a monotonically decreasing function, and therefore, the optimal bin size diverges. By contrast, the MISE of inhomogeneous point processes may have a minimum at some finite $\Delta$, provided that \begin{eqnarray}
\left. \frac{dS}{d(1/\Delta)}\right|_{\Delta=\infty} < 0.
\end{eqnarray}
This can be summed up as
\begin{eqnarray}
\frac{1}{\left\langle \lambda \right\rangle} \int_{-\infty}^{\infty} \left\langle \delta\lambda(t+s)\delta\lambda(t) \right\rangle ds > 1,
\label{detectable}
\end{eqnarray}
on condition that  $\int_0^{\infty}s \phi(s) ds$ is finite. Note that this instability condition derived from the histogram optimization is identical to the instability condition derived from the marginal likelihood maximization of the Bayesian rate estimator~\cite{Koyama07}.


Applying the above-mentioned consideration to the self-exciting process, we may obtain the condition for the detectable-undetectable criticality. The integral of the correlation function is given by the Fourier zero-mode, or, $\int_{-\infty}^{\infty} \left\langle \delta\lambda(t+s)\delta\lambda(t) \right\rangle ds = \int_{-\infty}^{\infty} \phi(s) ds = \tilde \phi_0$. For the self-exciting point process, the critical condition is obtained from Eq.(\ref{fouriercorrelation}) and $\tilde f_0=\int_{-\infty}^{\infty} f(t) dt =1$ as
\begin{eqnarray}
\frac{1}{\langle \lambda\rangle}\int_{-\infty}^{\infty}\langle \delta\lambda(t+s)\delta\lambda(t)\rangle ds = \frac{2\alpha-\alpha^2}{\left(\alpha-1\right)^2} = 1.
\end{eqnarray}
Thus the rate fluctuation in the self-exciting point process is detectable or undetectable, respectively if the excitability is larger or smaller than the critical value of
\begin{equation}
\alpha_c=1-1/\sqrt{2} \approx 0.2929.
\label{criticalexcitability}
\end{equation}
Note that this bursting transition occurs with the excitability much smaller than $\alpha=1$, at which the pandemic explosion occurs. Sample series of events generated with the excitability larger and smaller than the critical value are demonstrated in Fig.~\ref{figschematic}, from which we may observe burst of event occurrences and the apparent absence of rate fluctuation, respectively. 

\begin{figure}[ht]
\begin{center}
\includegraphics[clip,width=8.6cm]{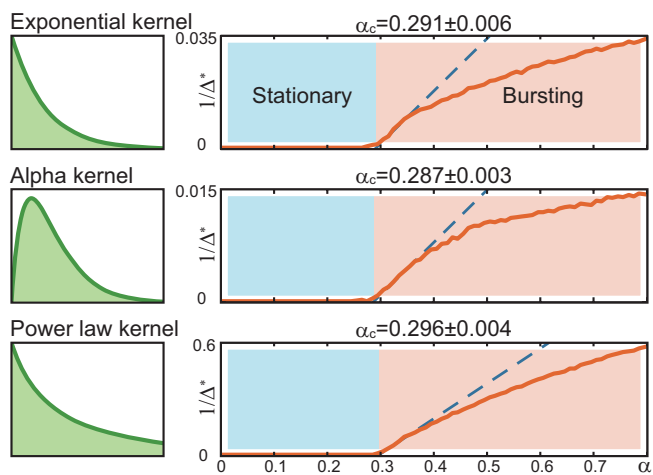}
\caption{{\bf Transitions in self-exciting processes of different kernels.} Inverse optimal bin size $1/\Delta^*$ is plotted with the excitability $\alpha$. Event sequences were generated with the exponential kernel, $f(t)= 0.1 \exp(-0.1 t)$ (top), the alpha kernel, $f(t)= 0.01 t \exp(-0.1 t)$ (middle), and the power law kernel, $f(t) = 0.5 (1+t)^{-3}$ (bottom). The bin size was selected by applying the $L^2$ optimization method~\cite{Shimazaki07} to sequences of $500,000$ events generated by the self-exciting point processes of $\rho=1$. The blue dashed lines represent regression lines fitted to $1/\Delta^*$ for an interval $\alpha \in [0.31,0.36]$.}
\label{figtransition}
\end{center}\end{figure}

The detectability of rate fluctuation can be quantitatively examined using principled rate estimation methods. We first generated event sequences of self-exciting processes with three kinds of kernels; the exponential function $f(t)= \tau^{-1} e^{-t/\tau}$, the alpha function $\tau^{-2} t e^{-t/\tau}$, and the power law function $(1+t)^{-3}/2$. For each series of events obtained under given excitability $\alpha$, we determined the optimal bin size $\Delta^*$ by using a method that enables to minimize the expected MISE even without knowing the underlying rate~\cite{Shimazaki07}. Figure~\ref{figtransition} shows how the inverse of the optimal bin size varies with the excitability. We observe that $1/\Delta^*$ begins to deviate from 0 when the excitability $\alpha$ exceeds some critical value. The critical excitabilities $\alpha_c$ estimated by the linear regression analysis are consistent with the theoretical value.

\begin{figure*}[ht]
\begin{center}
\includegraphics[clip,width=\linewidth]{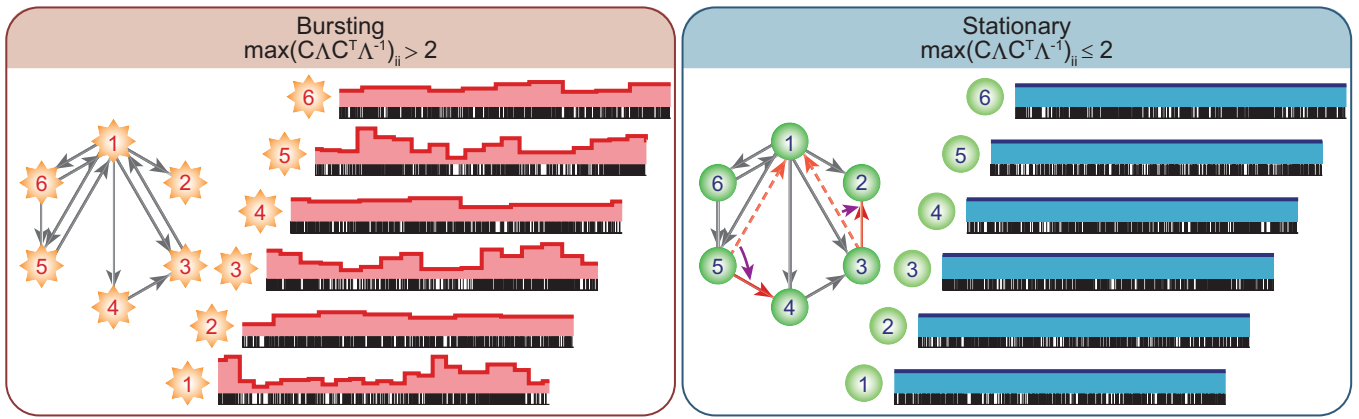}
\caption{{\bf Multi-dimensional self-exciting processes.} Networks of $N=6$ agents linked with $10$ connections of the excitability $\alpha_{ij}=0.4$ with the kernel $f(t)= 0.2 \exp(-0.2 t)$. (a) A network exhibits bursts of event occurrences. (b) Bursting is inhibited by reconnecting agents (from dashed lines to solid lines).}
\label{fignetwork}
\end{center}\end{figure*}

Finally, we extend the theory to multi-dimensional self-exciting processes to discuss the criticality in networks of agents, such as people communicating with emails. Let $\lambda_i(t)$ and $\nu_i(t)$ represent the occurrence rate and a series of events of the $i$th agent ($i=1,2,\cdots,N$). The multi-dimensional process is given by
\begin{equation}
\lambda_i(t)=\rho_i+\sum_{j=1}^N \alpha_{ij} \int_{-\infty}^{\infty} f(t-s) \nu_j(s)ds,
\end{equation}
where $\rho_i$ is the base rate and $\alpha_{ij}$ represents the inter-agent excitability or the supplementary probability for $i$th agent caused by an event of $j$th agent.

Given an excitability matrix $\bm{\alpha} = \{\alpha_{ij}\}$, the average firing rate $\bm{\langle \lambda \rangle} = \{\langle \lambda \rangle_i\}$ is obtained from $\bm{\rho} = \{\rho_i\}$ as
\begin{equation}
\bm{\langle \lambda \rangle}=\bm{C}\cdot\bm{\rho},
\end{equation}
where $\bm{C}$ represents effective connections~\cite{Rotter13},
\begin{equation}
\bm{C} \equiv \sum_{n=0}^{\infty}\bm{\alpha}^n = \left(\bm{I}-\bm{\alpha}\right)^{-1}.
\end{equation}
Thus all eigenvalues of $\bm{\alpha}$ should be smaller than $1$ to avoid pandemic explosion. 

We represent the correlation functions of the rate fluctuation of agents or event sources by a matrix $\bm{\phi}(s)$. Similarly to the one-dimensional process (\ref{fouriercorrelation}), we may obtain the Fourier image of the correlation matrix $\tilde{\bm\phi}_{\omega}$ \cite{Hawkes71b}. In particular, we may obtain the Fourier zero-mode as $\tilde{\bm{\phi}}_0 = \bm{C} \bm{\Lambda}\bm{C}^T-\bm{\Lambda$}, where $\bm{\Lambda} = {\rm diag}\left(\bm{\langle \lambda \rangle}\right)$. From this, we can obtain the condition under which a network of agents exhibits fluctuating rate. That is
\begin{equation}
\max_i \left(\bm{C}\bm\Lambda\bm{C}^T\bm\Lambda^{-1}\right)_{ii} > 2.
\label{ConditionOfMutual}
\end{equation}

It is possible to incite or inhibit the bursting by changing the connections among agents by taking account of the above condition. We demonstrated this by simulating a network of $N=6$ agents linked with $10$ connections. The network satisfying the condition (\ref{ConditionOfMutual}) exhibits bursts of event occurrences, whereas we are able to eliminate the bursting by reconnecting the agents (Fig.~\ref{fignetwork}).


In this study, we have shown that a linear self-exciting point process undergoes a transition at which the rate fluctuation changes from ``invisible" to ``visible". It should be noted that this transition does not belong to typical critical phenomena: Because the event generation process is linear, the correlation function varies smoothly with the excitability. The criticality is derived from observers in such a way that the rate fluctuation becomes ``visible" in the sense of $L^2$ measure. In deriving this criticality condition, we assumed a second-order phase transition in which $1/\Delta^*$ continuously deviates from $0$. It is possible that the occurrence rate destabilizes from a finite time scale under the smaller excitability, but in practice this occurs only if the kernel possesses a very strong oscillation component of a finite time scale. We further extended the analysis to multi-dimensional processes and derived the necessary condition for bursting. Knowing the criticality condition, Eq.(\ref{ConditionOfMutual}), we can control bursts of events in agent networks by reorganizing the connectivity among the agents.

\section*{ACKNOWLEDGMENTS}

We thank Rob Kass and Shuhei Kurita for stimulating discussions. This study was supported in part by Grants-in-Aid for Scientific Research to SS from the MEXT Japan (25115718, 25240021), and by JST, CREST.

\end{document}